Stresa, Italy, 25-27 April 2007

# NANOBIOSENSORS BASED ON INDIVIDUAL OLFACTORY RECEPTORS

*Edith Pajot-Augy*

Institut National de la Recherche Agronomique, UMR1197 Neurobiologie de l'Olfaction et de la Prise Alimentaire (NOPA), Récepteurs et Communication Chimique (RCC), F-78350 Jouy-en-Josas, France

**ABSTRACT**

In the framework of the SPOT-NOSED European project, nanoscale sensing elements bearing olfactory receptors and grafted onto functionalized gold substrates are used as odorant detectors to develop a new concept of nanobioelectronic nose, through sensitive impedancemetric measurement of single receptor conformational change upon odorant ligand binding, with a better specificity and lower detection thresholds than traditional physical sensors.

## 1. INTRODUCTION

Animals such as rats, dogs, but also bees, can detect Volatile Organic Compounds specifically and with very low thresholds, which is used in fields as various as explosives or drugs detection, and also for medical diagnosis in cases where VOCs are associated to a given pathology. Presently, electronic noses are based on the detection and discrimination of complex odorant mixtures via a network of sensors with a broad spectrum, that generate a characteristic response pattern. Comparing this profile to a database of patterns generated by standards allows the identification and qualification of the odorants present in the mixture. These chemical sensors rely on various physical techniques (electrical, optical, calorimetric, gravimetric, ...) to estimate parameters able to discriminate the odorants present. However, if those sensors display some sensitivity and capacity to analyze odorant mixtures, objectivity, and rapidity, they are nevertheless limited in terms of specificity, stability, and hampered by the presence of water. Electronic noses are presently developed and miniaturized at the level of micro e-noses. Meanwhile, a number of cell-based biosensors have been designed, using yeasts, or more specialized taste or olfactory sensory neurons, or even cells recombinantly expressing olfactory receptors, so as to measure changes of electrical parameters in the presence of odorants.

A further step is to use the olfactory receptors themselves as a source of sensing elements for biosensors, using various methods such as BioFETs, SPR, piezoelectricity, or 2nd messengers, to monitor their functional response to odorants. Such sensors benefit the natural biological optimization of the molecular recognition of an odorant by its receptor, which ensures the response specificity and reproducibility. They give a potential access to all kinds of odorants, with the very low detection thresholds observed on the animals ($10^{-9}$M...$10^{-12}$M...$10^{-14}$M), and they can work in an aqueous environment, contrary to chemical sensors. However, since receptors are expressed in low amounts in olfactory sensory neurons, and several hundred different receptors are expressed each in another neuron, the primary requirement is that of an adequate expression system for olfactory receptors. The yeast *S. cerevisiae* proved efficient to reach a quantitative functional expression and adequate trafficking of olfactory receptors to the plasmic membrane under optimized induction conditions.

## 2. SPOT-NOSED

The purpose of the European project "SPOT-NOSED" (http://www.nanobiolab.pcb.ub.es/projectes/spotnosed/) was the development of a nanobiosensor array based on the electrical properties of single olfactory receptors, to mimic the performances of natural olfactory sensing system.

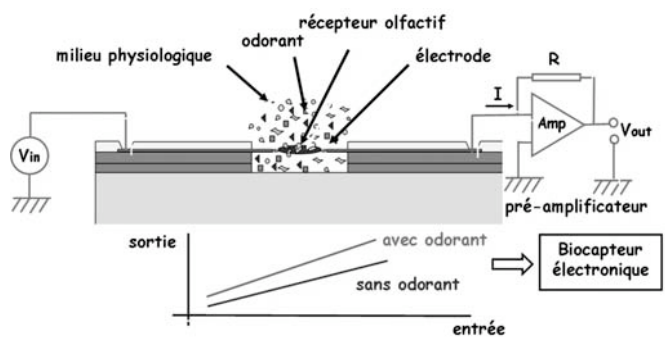

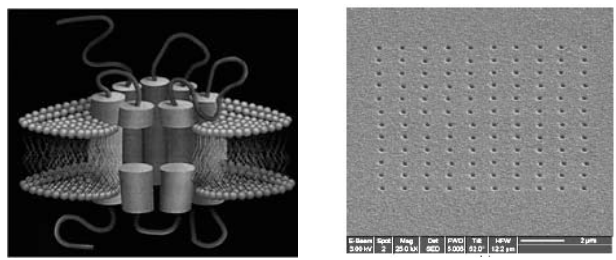

Figure 1 : Principle of *in vitro* odorant detection and identification. A schematic olfactory receptor is shown (7 transmembrane G protein coupled receptor), and a possible aspect for a sensor array.

The nanobiosensor array will integrate nanotransducers consisting of gold nanoelectrodes with a "single" olfactory receptor anchored in it.

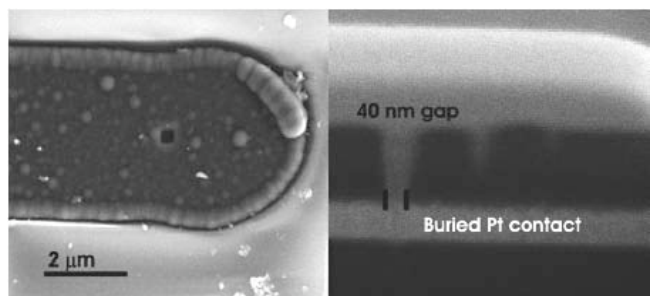

Figure 2 : Scanning electron microscope image of a nanoelectrode fabricated by focused ion beam milling and cross section showing an aperture of 40 nm [1].

### 3. NANOSOMES FROM SNIFFING YEASTS

In the framework of this project, an OR and an appropriate $G_\alpha$ protein were co-expressed in yeast cells [2] from which membrane nanosomes, with size homogenized to around 50 nm by sample sonication, were prepared.

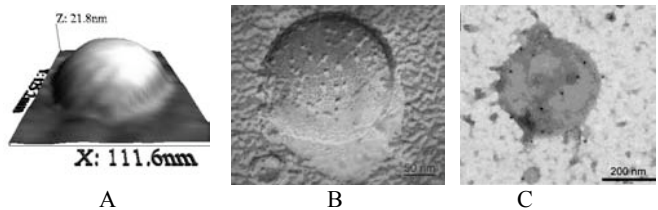

Figure 3 :
A - 3D AFM topographic image of a single non-sonicated nanovesicle adsorbed on thiol functionalized gold. Image taken in PBS environment using Jumping mode [3].
B - 200 x 200 nm transmission electron microscopy image of a nanosome prepared by cryo-fracture. Membrane proteins are clearly observed [4].
C - Electron microscopy of an immunogold-labeled nanosome : gold grains indicate the presence of the receptor [5].

The nanovesicles containing functional olfactory receptors are grafted onto bare or functionalized substrates without disruption, and provide a high surface coverage, as visualized by means of Atomic Force Microscopy.

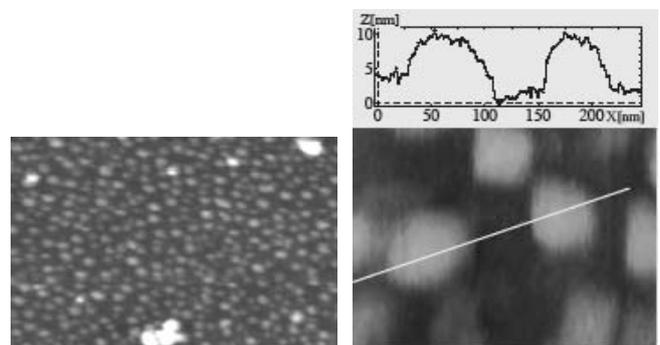

Figure 4 : AFM image of sonicated nanosomes on bare gold (coverage 50 %)[4].

### 4. SURFACE FUNCTIONALIZATION AND FUNCTIONAL TESTS

Gold surfaces were sequentially functionalized with mixed Self Assembled Monolayers to provide a controlled density of anchoring sites, neutravidin, and specific biotinylated receptor antibody. This also allows a proper localization and orientation of receptors on the surfaces, and endows stability to the assembly.

The efficiency of transfer and specificity of immobilization were checked with several techniques : Surface Plasmon Resonance (SPR) [5, 6], Electrochemical Impedance Spectroscopy (EIS)[7, 8], Quartz Crystal Microbalance (QCM) [9], and AFM [4, 5].

Several procedures were compared to attach thiols to the gold chip, and two modes of nanosomes immobilization were evaluated : nonspecific "adsorption" onto SAM, and "capture" by grafting via a specific antibody to the receptors, using SPR.

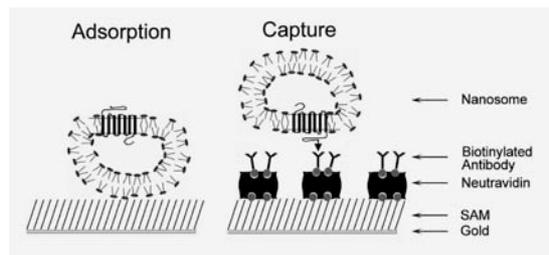

Figure 5 : Specific immobilization of nanosomes on gold surface functionalized by a mixed Self Assembled Monolayer [6]. Schematic view of the two modes of nanosomes immobilization: nonspecific "adsorption" onto SAM, and "capture" by grafting via a specific antibody [5].

Functional tests consist in recording the SPR signal shift due to $G\alpha$ release upon odorant stimulation in the presence of GTPγS.

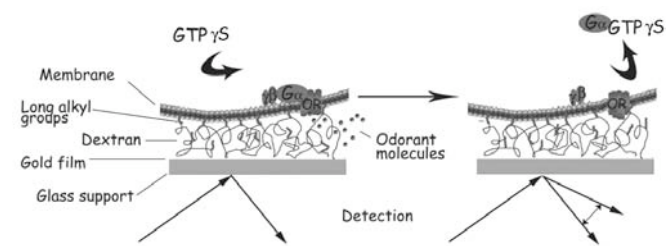

Figure 6 : Principle of the SPR sensor system. A L1 sensor chip (BIAcore) is composed of a gold film deposited onto a glass support, on which a carboxylated dextran grafted with long alkyl groups is attached. Nanosomes are immobilized via interaction of their lipidic surface with these long alkyl chains, leaving membrane receptors fully available for ligands and G protein interactions. The scheme presents only a restricted area of an intact nanosome. Upon stimulation of an OR with an odorant ligand, the $G\alpha$ subunit is activated and desorbed in the presence of GTPγS. This results in a shift of the SPR response level expressed as arbitrary resonance units (RU) [10].

This test has previously demonstrated that ORs carried by isolated nanosomes immobilized on sensorchips retain their full activity and discriminate between odorant ligand and unrelated odorants, as previously shown in whole yeast cells with a reporter gene.

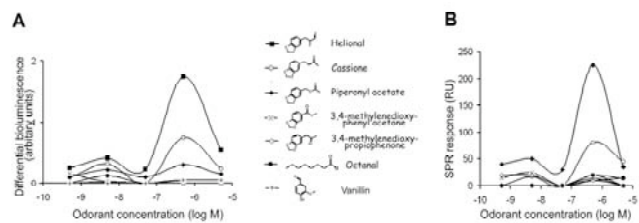

Figure 7 : Detection of odorants by cmyc-OR1740 in whole yeast cells (A) and by the receptor in immobilized nanosomes (B).
A - Differential bioluminescence responses upon stimulation of the receptor in whole yeast with the same panel of odorants. In both cases, concentration-dependent curves are plotted as a difference of response to odorants relative to controls obtained by replacing odorants with water.
B - Differential SPR responses obtained upon stimulation of the ORs in nanosomes with various odorants in the presence of 10 μM GTPγS. Odorants tested are shown in the middle panel [10].

Here, the SPR experiments demonstrated that the intensity of the specific response was enhanced in the case of nanosomes captured via an antibody to the receptor compared to adsorbed nanosomes.

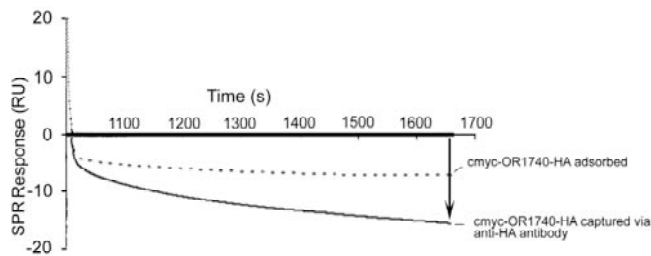

Figure 8 : Differential SPR sensorgram showing signal modifications observed when helional and GTPγS are simultaneously injected over immobilized nanosomes. Signal responses are normalized to the corresponding controls obtained by replacing odorant with water. The intensity of the specific response is enhanced in the case of nanosomes captured via an antibody to the tag (arrows) compared to adsorption [5].

## 5. MICROCONTACT PRINTING

Microcontact printing was then used to immobilize SAMs onto gold surfaces. Positive and negative elastomeric polydimethylsiloxane stamp were replicated from silicon-based molds elaborated using deep reactive ion etching. Positive stamps present round posts of 2.5 μm diameter that protrude 800 nm from the surface. Patterning was performed by inking procedures of the stamp after hydrophilization with oxygen plasma in order to ensure correct surface coverage.

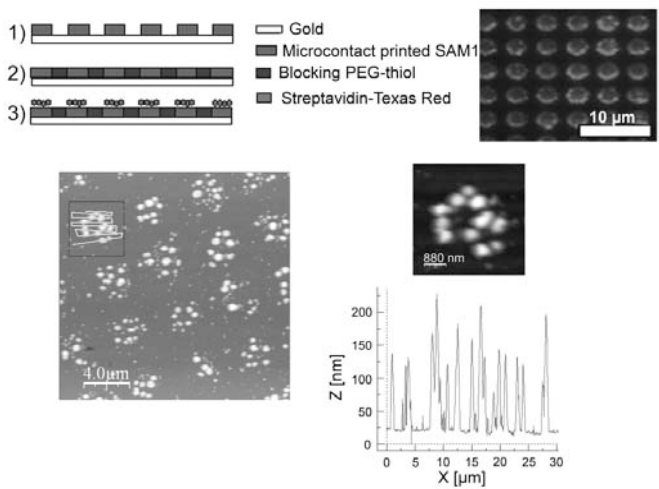

Figure 9 : Schematic representation of the procedure used to immobilize SAM onto the gold surfaces by microcontact printing. The patterning of surfaces is confirmed by fluorescent images of streptavidin labeled with a fluorescent dye, attached onto SAM biotinyl groups. An AFM image shows features at the nanometric scale on the surface functionalized with SAM after nanosomes deposition. High-resolution topography of the area outlined by the box is shown, and a horizontal cross section along the developed green line from reveals the thickness of the immobilized features [5].

Electrochemical Impedancemetric Spectroscopy can also be performed at each step of surface functionalization and nanosomes grafting, and significant changes were detected upon the final grafting of increasing receptor amounts.

Modelization of single receptor electrical behavior, based on the only appropriate prototype (rhodopsin, only known GPCR 3D structure) allows the equivalent circuit modelling of an OR, providing a corresponding impedance network model.

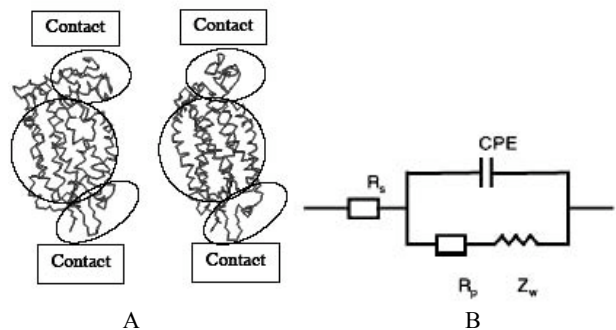

Figure 10 : Modelization of single receptor electrical behavior [11].
A - Rhodopsin fundamental and activated states
B - Corresponding impedance network model

The conformational change induced by ligand binding is predicted to yield a global impedance change up to 50%. Dedicated electronic instrumentation was designed to produce high-gain, very low-noise amplifiers with extended bandwidth, to measure the nanobiosensors electrical response with Scanning Probe Microscopy. A user friendly interface with specific odorant identification algorithm was developed. The own electrical properties of individual olfactory receptors are thus expected to endow these impedancemetric sensors with a far better specificity, detection threshold, reproducibility, and odor spectrum than traditional physical sensors.

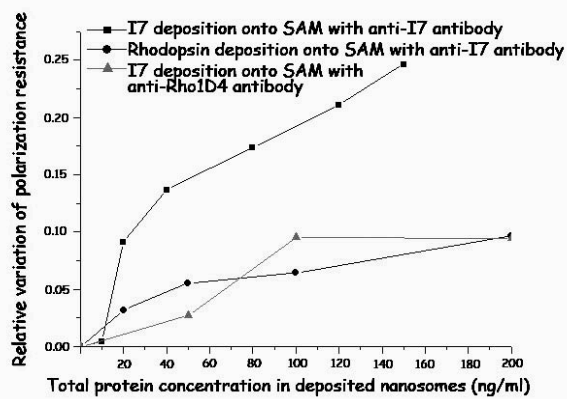

Figure 11 : Electrochemical Impedance Spectroscopy
A - Deposition of nanosomes onto various antibodies immobilized on gold electrodes : the immobilization of nanosomes is specific [3]

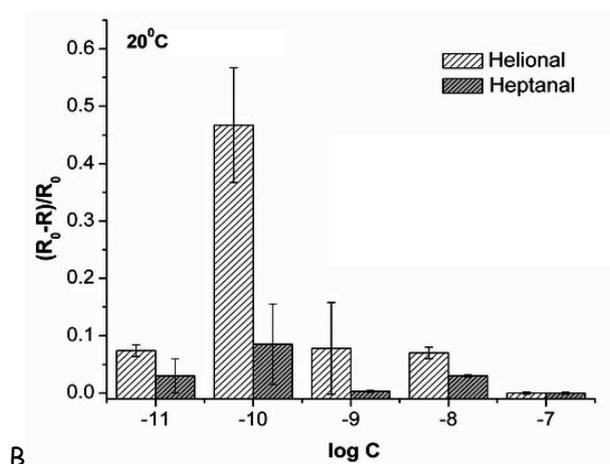

B - Functional response of ORs in immobilized nanosomes. EIS response of cmyc-OR1740 on immobilized nanosomes to helional is specific and concentration-dependent : immobilized olfactory receptors discriminate between odorants [12]

These findings will allow the development of a new generation of nanosensors, bioelectronic olfaction devices for rapid and noninvasive assessment of VOCs, that can constitute a signature of metabolic states or diseases, participate in aromas in food, be associated with drugs and explosives or to domestic and environmental pollutants. Ligand screening for orphan receptors and pharmacological screening may constitute further high value applications.